\documentclass[12pt,preprint]{aastex}

\slugcomment{ApJL in press}

\shorttitle{Diameters and albedos of three Near Earth Objects}
\shortauthors{Trilling et al.}

\begin{document}


\title{Diameters and albedos of three sub-kilometer Near Earth Objects derived
from Spitzer observations}


\author{D. E. Trilling$^{1}$,
M. Mueller$^{1}$,
J. L. Hora$^{2}$,
G. Fazio$^{2}$,
T. Spahr$^{2}$,
J. A. Stansberry$^{1}$,
H. A. Smith$^{2}$,
S. R. Chesley$^{3}$, \&
A. K. Mainzer$^{3}$}

\affil{
1: Steward Observatory, 933 N. Cherry Avenue,
The University of Arizona, Tucson, AZ 85721;
{\tt trilling@as.arizona.edu}}
\affil{
2: 
Harvard-Smithsonian Center for Astrophysics,
60 Garden Street, 
Cambridge, MA 02138}
\affil{
3: Jet Propulsion Laboratory,
Caltech, 4800 Oak Grove Drive, Pasadena, CA 91109}








\begin{abstract}
Near Earth Objects (NEOs) are fragments of
remnant primitive bodies
that date from the era of Solar System formation.
At present, the physical properties and origins of NEOs are poorly
understood.  
We have measured
thermal emission from 
three NEOs --- (6037) 1988~EG,
1993~GD, and 2005~GL ---
with Spitzer's IRAC instrument
at 3.6, 4.5, 5.8, and 8.0~$\mu$m (the last object
was detected only at 5.8~and 8.0~$\mu$m).
The diameters of these three objects are
400~m, 180~m, and 160~m, respectively, 
with uncertainties of around 20\% 
(including both observational and systematic
errors).
For all three the
geometric
albedos are
around~0.30, in agreement with previous results 
that most NEOs
are 
S-class asteroids.
For the two objects detected at 3.6~and 4.5~$\mu$m,
diameters and albedos based only on those data agree with the values
based
on modeling the data in all four bands. This agreement, and the high
sensitivity
of IRAC, show the promise of the Spitzer Warm Mission for determining the
physical
parameters for a large number of NEOs.
\end{abstract}





\section{Introduction}

Near Earth Objects (NEOs)
are
bodies whose orbits pass within a few tenths of an AU of the Earth's orbit. As of this
writing, there are around 5000 NEOs known. The Pan-STARRS program is
likely to increase the number of known NEOs to $\sim$10,000 or more by 2013.
These bodies are of critical interest to both the scientific
community and the public.
The NEO population is the source of potential
Earth-impacting asteroids (hence a Congressional mandate
to study these objects), 
and some may be
easily reached by spacecraft, enabling our exploration of the nearby Solar
System.
Because NEOs have only recently been perturbed out of orbits in the
main asteroid belt, and so are relatively primitive objects, they contain
information that records the origin
of our Solar
System and that may offer insight into both the past (via delivery of
organic material) and
future (via impact-caused extinctions) of life on Earth.
However, the physical characterization of these objects is by far outpaced by discoveries.
The NEO size and albedo distributions, crucial inputs for
Solar System studies as well as the assessment of the NEO Earth impact hazard, 
are only poorly constrained, especially at the smallest sizes \citep[e.g.,][]{StuartBinzel2004}.


The diameter and albedo of asteroids can be determined from thermal-infrared observations together with appropriate thermal modeling 
\citep[e.g.,][]{STM,LebofskySpencer1989,HarrisLagerros2002}, provided the absolute magnitude $H$ 
\citep[optical magnitude at a standardized observing geometry; see][]{HG}
is known.
A suitable thermal model for NEOs is the
Near-Earth Asteroid Thermal model \citep[NEATM, ][]{neatm}, which
allows for simultaneous fits of the
asteroid diameter, albedo, and effective surface temperature (parametrized through the beaming parameter $\eta$)
\citep[e.g.,][]{neatm,HarrisLagerros2002}.


We have measured
thermal emission from
three NEOs with the Spitzer Space
Telescope \citep{werner04}
and present our data (\S2) and results (\S3) here.
Using the NEATM, we derive albedos and diameters for all three objects (\S3).
In \S4 we comment on the apparent thermal inertias for these objects
and 
demonstrate that a study of NEOs could profitably
be carried out with the Spitzer Warm Mission.

\section{Observations and data reduction}

We observed three NEOs ((6037) 1988~EG;
1993~GD; and 2005~GL) at 3.6, 4.5, 5.8, and 8.0~$\mu$m
with Spitzer's InfraRed Array Camera \citep[IRAC;][]{fazio04} using the moving cluster
observing mode, tracking according to the standard NAIF ephemeris.
Table~\ref{geom} gives
the observing log and observing geometries.
These objects were chosen to be visible by Spitzer 
on our observing date, have small positional uncertainties,
and have a range of absolute magnitudes.
IRAC observes simultaneously at
[3.6, 5.8]~$\mu$m and at
[4.5, 8.0]~$\mu$m.
Our dithered observations alternated
between these two pairs of bandpasses
to reduce the
relative effects of any lightcurve variations within the observing period, and
to maximize the relative motion of the asteroid to help reject
background sources.
The high dynamic range mode was
used, and the frame times for each object are given in Table \ref{geom}.

We used the Basic Calibrated Data (v.\ 17.2.0) and MOPEX (v.\
16.0) in the moving object mode to construct mosaics of the
fields in the reference frame of the asteroid, assuming the offsets given
by the NAIF ephemeris.  Aperture photometry was
performed using an aperture radius of 5 pixels (at 1.22 arcsec/pixel) and an
annulus of 5 pixels for background measurement around the source.  The aperture
and annulus sizes were calibrated using one of the IRAC calibration stars
(HD 165459) and the zero point set so that the measurements matched the
source magnitudes given in \citet{reach05}.
6037 and 1993~GD were detected in all four bands;
2005~GL was only detected at 5.8~and 8.0~$\mu$m (Table~\ref{geom}).
Only 6037 is bright enough to provide good
time-series photometry; no significant flux
variation was detected in the $\sim$15~minute span
of the observations.


Due to the spectral width of the IRAC passbands, measured flux values must be color corrected.
The observed asteroid fluxes comprise 
thermal and reflected light components, which have different
color corrections (though color correction for the
latter is negligible).
To derive color-corrected thermal fluxes, 
we must first remove the reflected flux contribution in each band.
The flux component from reflected sunlight was assumed to have the spectral shape of a $T=5800$~K black body over IRAC's spectral range.  The flux level was determined from the solar flux at 3.6~$\mu$m \citep[$5.54\times10^{16}$~mJy, ][]{solarspec}, the solar magnitude of $V=-26.74$, and the asteroid's V magnitude as determined from the observing geometry and the known $H$ value.
We assumed that asteroid reflectivity at 3.6~$\mu$m and longward is 1.4~times the reflectivity
in the V~band (A. Rivkin, pers.\ comm.), although using the naive assumption
of equal reflectivity makes only a few percent difference in the resulting
albedos and diameters.
The estimated reflected light component was subtracted
from the measured fluxes to get the (uncorrected) thermal fluxes.

Color-correction factors for the thermal flux were determined using the iterative procedure described in \citet{Mueller2007}: Color-correction factors were first determined for typical NEATM parameters,
the resulting fluxes were fit using the NEATM, then color-correction factors were re-derived using the best-fit NEATM parameters until convergence was reached.
Color corrections and
color-corrected thermal fluxes for all three targets are given in Table~\ref{results}.

\section{Model results and uncertainties}

\begin{figure}
\begin{center}
\includegraphics[angle=270,scale=0.50]{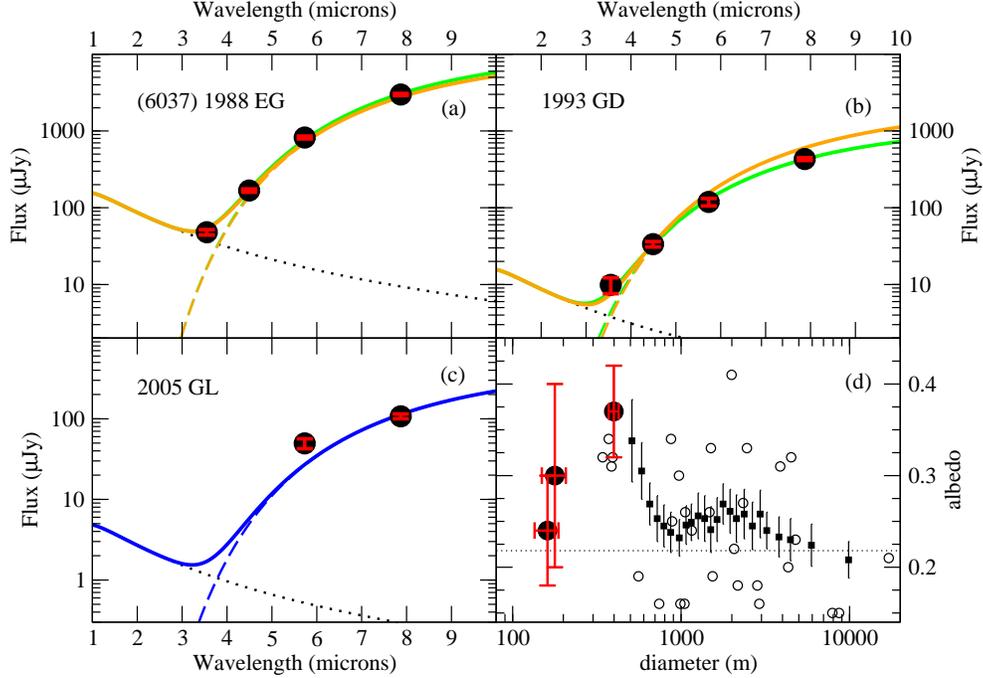}
\end{center}
\caption{{\em Panels a,b,c:} Spectral energy distributions for
the three observed NEOs.
Data points (black filled symbols with
red error bars overplotted) show
our color-corrected total (reflected plus
thermal) fluxes. 
Green curves (panels a,b) show fits to four band
data. Orange curves (panels a,b) show
fits to data at 3.6~and 4.5~$\mu$m only, with
$\eta$ fixed.
Blue curves (panel c) show fits to
data at 5.8~and 8.0~$\mu$m only,
with $\eta$ fixed.
Dashed lines indicate thermal components and
dotted lines indicate reflected
light components of the total flux, which
is plotted with solid lines.
In panels (a) and (b) the orange curves lie
nearly on top of the green curves, implying that
the two fits are very similar (but making the
green curves difficult to see).
{\em Panel d:} A modified version of
Figure~8 from \citet{binzel04} that
also plots the average diameters and albedos that
we report here as large black circles with
red error bars; see Table~\ref{results}.
The open small black circles are individual data points
and
filled black squares are mean values for
S-class (and related classes) NEOs \citep{binzel04}.
The dotted line is the mean
albedo for main belt S class asteroids.
The error bars on the solutions here reflect the
scatter in the model solutions, but do not include
additional uncertainties that may derive from ignorance
of physical properties of the asteroids or
uncertainties in $H$ (see text for discussion).
\label{sed}}
\end{figure}

Thermal fluxes were measured in all four IRAC bands for
(6037) 1988~EG and 1993~GD.
For each target, the four-band data were fit using the NEATM
by varying diameter $D$, albedo $p_V$, and $\eta$ until $\chi^2$ was minimized. $D$ and $p_V$ are related through the optical magnitude $H$: $p_V = 10^{-H/2.5} \left( 1329~\textrm{km}/D\right)^2$ \citep{FowlerChillemi}. (In all cases we assume
emissivity of~0.9 and standard scattering behavior in the visible,
resulting in a phase integral of~0.39.) 
These best-fit (floating $\eta$) values for $D$, $p_V$, and $\eta$ are given in Table~\ref{results} and
the corresponding model spectra are shown
in Figure~\ref{sed}.

For these four-band (floating $\eta$) fits, we use a Monte Carlo analysis
to estimate the statistical uncertainty of our results.
300~random sets of flux values were generated such that their mean and     
standard deviation match the measured fluxes and flux uncertainties, respectively.        
Each trial was fit using the NEATM as above.
The standard deviations of the resulting 
diameter and albedo values were taken to be
the statistical uncertainties on our four-band results (Table~\ref{results}).
We do the same NEATM/Monte Carlo analysis using just
the 5.8~and 8.0~$\mu$m data for the brighter two objects (Table~\ref{results}),
allowing us to assess systematic variations in model
results.
However, 
because the measurements of 2005~GL have relatively
low significance, this Monte Carlo approach
does not work, and 
we require a proxy technique to determine
variations among models, as follows.

With this Monte Carlo proxy model, the
nominal fit is determined
in the usual way (NEATM, as above). We then
use the NEATM
to fit a ``hot'' solution, where the short wavelength
data is increased by 0.7$\sigma$ and the long wavelength
solution is decreased by 0.7$\sigma$ relative to
the nominal flux values ($\sigma$ is the
measurement error). We also fit a ``cold''
solution, which has short band decreased and the 
long band increased by 0.7$\sigma$.
The range in derived albedo
and diameter then is derived from the range of values
produced by the hot and cold solutions.
We show in Table~\ref{results}
that for 6037 this hot/cold proxy approach replicates
the full Monte Carlo result quite closely.
We present this proxy model here because,
in general, this approach is a useful substitute
for full Monte Carlo modeling.
However, for 2005~GL, the
significance of our measurements is so poor that even this technique
does not work (producing implausible albedos
around~2 and unlikely $\eta$~values around~0.38).
We must therefore move to yet a 
simpler technique to assess
systematic errors due to model variations.

\citet{Delbo2003,Delbo2007,wolters} derived an empirical relationship
between the phase angle $\alpha$ at which observations
are made and the best-fit $\eta$.
This ``fixed $\eta$'' technique works here
because the number of free parameters is decreased
by one: the surface temperature is fixed
due to the fixed $\eta$ (compare to the hot/cold
models above).
Thus, we produce ``fixed $\eta$ from $\alpha$''
solutions for 2005~GL, as well as for 
6037 (fitting 5.8~and 8.0~$\mu$m data and fitting
3.6~and 4.5~$\mu$m data) and for 
1993~GD (with the same data subsets) (Table~\ref{results}).
In the interest of assessing variations due
to different model solutions, we also derive
``fixed $\eta$ from $\alpha$'' solutions
for 2005~GL using just 5.8~$\mu$m data and
using just 8.0~$\mu$m data (Table~\ref{results}).
The formal errors
on these fixed $\eta$ solutions are derived
directly from the measurement errors:
because any acceptable fit must pass within 
the measurement error bars, the percent
error on diameter is equal to the percent
error on the best measurement utilized in the fit,
divided by two (since flux is proportional to
diameter squared).
The albedo uncertainty is twice that of the 
diameter uncertainty, or equal to the uncertainty
on the best measurement utilized.

We take our 
final model solutions to be the averages
of the solutions from the various techniques (Table~\ref{results}).
This allows us to capture the scatter among
the different model solutions
in the error bars on our final solutions.
The uncertainties on diameter
are around 7\% for the strongly 
detected 6037 and around 16\% for the less well 
detected 1993~GD.
For 2005~GL, where there are only three models, all
of the same type, the uncertainty on diameter
is also around 16\%.
Uncertainties on albedo are twice those on
diameter.
These final solutions are given in 
Table~\ref{results} and plotted in
Figure~\ref{sed}.


%
%
%
%

Our diameter solutions are hindered by our
lack of knowledge about physical target properties such as shape, spin state, thermal inertia, and surface roughness, all of which affect surface temperatures and hence thermal flux; together, these typically
imply an uncertainty around 15\% \citep[e.g.,][]{Wright2007},
comparable to the systematic errors
we estimate from our cross-model comparisons above.
More realistic thermophysical modeling \citep[e.g.,][]{HarrisLagerros2002} would require models for shape and spin state as inputs, but those are unlikely to become available for our targets in the near future.

Additional diameter uncertainty derives from the rotational flux variability (lightcurves) of our       
targets.
The peak-to-peak lightcurve amplitude of 6037 is\footnote{{\tt http://www.asu.cas.cz/$\sim$ppravec/neo.html}} around 0.2~mag.
Following the arguments
presented in Appendix~A, we find that the resulting diameter uncertainty due to 
lightcurve effects is
negligible: less than 4\%.
Nothing is known about the lightcurves of our remaining targets. Given their small
size, their lightcurves are likely to have a rather large amplitude and small period
\citep{pravec}.
By virtue of our observation design, measured fluxes in all four channels are
effectively averaged over $\sim$900~s (1993~GD) and $\sim$2,000~s
(2005~GL).
Assuming a lightcurve amplitude of 1~mag and a period of 1 hour for 2005~GL, the
corresponding diameter uncertainty due to lightcurve effects would be around 2\% -- negligibly small.

The final diameter uncertainties are therefore the combination 
of uncertainties from modeling ($<$20\%);
uncertainties in physical properties (15\%);
and lightcurve effects (small). The total uncertainties
on diameters are likely to be around 20\%, including
errors from both measurement and systematic
uncertainties.


The corresponding albedo uncertainty due to 
scatter in model results and ignorance of 
physical properties 
is therefore around 40\%.
Uncertainties in $H$, which leave the best-fit diameter estimate practically unchanged \citep{HarrisHarris}, add to the error budget for $p_V$. 
This effect is small for 6037 (where the uncertainty
in $H$ is estimated to be 0.15~mag), but $H$ could be in error by 0.3~mag or more for the
other two targets, leading to errors in $p_V$ of 30\% or more.
Combining these two uncertainties (40\% from above
and 30\% from $H$ uncertainty), we therefore
estimate the total uncertainty
on our albedo determinations to be around 50\%.

\section{Discussion}

All three objects have diameters less than 500~meters, making them
among the smallest NEOs with known albedos and diameters, and 
among the smallest individual objects studied with the Spitzer
Space Telescope. All three objects also have albedos close
to~0.3,
in agreement with the idea that the NEO
population is dominated by S-class asteroids
\citep[e.g.,][]{binzel04}.
\citet{binzel04} also found that 
the albedos for S-class (and related classes)
NEOs rise from their main belt average value
around~0.22 to greater than~0.3 for objects
$\lesssim$500~m. 
Our results appear to confirm
this trend (Figure~\ref{sed}),
though with small numbers and 
not insignificant error bars.
It is quite premature to discuss
the reality of the potentially interesting downward turn
at even smaller sizes.

The best-fit (floating) $\eta$ values found for 6037 and 1993~GD are roughly consistent with empirical expectations \citep{Delbo2003}, which were recently used by \citet{Delbo2007} to determine the typical thermal inertia of $D\sim 1$~km NEAs.
Thermal inertia is indicative of the presence or absence of loose material (regolith) on the surface and is a key parameter for model calculations of the Yarkovsky effect, a non-gravitational force that severely influences the orbital dynamics of small asteroids.
(Note that \citet{Vokrouhlicky2005} list 6037 as a potential target for direct observations of the Yarkovsky effect.)
Our results suggest that our targets have unremarkable thermal inertias
and may be similar to the 
320~meter diameter S-type NEO (25143) Itokawa \citep{ThMueller,MuellerDiss}, the target of the Hayabusa mission.
However, more work and a systematic, large survey 
are needed
to determine the typical thermal inertia of sub-km NEAs.


For 6037 and 1993~GD
the diameters and albedos we derive
using only 3.6~and 4.5~$\mu$m data
are in agreement with our other model
solutions, particularly for 6037,
which is strongly detected (SNR$>$10)
in both bands.
This agreement
has important implications for the Spitzer Warm Mission.
After Spitzer's onboard cryogen is exhausted, observations
in IRAC bands 1 and 2 (3.6 and 4.5~$\mu$m) can still be made
with essentially no loss of sensitivity. 
Our results
show the promise of capitalizing on the superior
sensitivity of IRAC
to determine the physical properties of 
a large number of NEOS during the Spitzer
Warm Mission.




\acknowledgments

We thank the referee for a number of useful suggestions.
We thank Tom Soifer for allocating Director's Discretionary
time for this project and Mike Werner
for helpful suggestions. We acknowledge the extremely rapid release
of these data by the SSC and
Sean Carey for MOPEX advice.
Andy Rivkin helped us estimate the relative spectral
reflectances of asteroids and 
Rick Binzel provided us his data
that we plotted in Figure~1.
This work is based on observations made with the Spitzer Space Telescope, which is operated by JPL/Caltech under a contract with NASA. Support for this work was provided by NASA through an award issued by JPL/Caltech.



Facilities: \facility{Spitzer(IRAC)}



\appendix

\section{The effect of unknown lightcurve variations on diameter uncertainties}

Uncertainties in diameter can arise from the rotational flux variability (lightcurve)
of an observed asteroid.
To first order, the projected area $A$ of an asteroid with
a double-peaked lightcurve varies as 
%
$A (\phi)/A_0 = 1+(10^{\frac{\Delta m/2}{2.5}}-1)\sin 2\phi$,
%
where $\phi$ is rotational phase, $A_0$ is the average area,
and $\Delta m$ is the peak-to-peak lightcurve
amplitude.
For an instantaneous area measurement at a random time, the expectation value 
is $A_0$ and the 
standard deviation is
%
$\sigma_A = A_0 (10^{\Delta m/5}-1)/\sqrt{2}$.
%
Since area is proportional to diameter squared, the
lightcurve-induced contribution to the fractional diameter
uncertainty is
%
$\sigma_{D} = (10^{\Delta m/5}-1)/\sqrt{8}$.
%
Therefore,
for objects whose lightcurve amplitudes but not
periods are known, $\sigma_{D}$ can
be estimated.
Only for objects with $\Delta m \geq 1.9$, which is a very large
amplitude lightcurve, is
$\sigma_{D}$ greater than 50\%.

Some observations may span a significant portion of
an asteroid's rotation period; our relatively
long integrations on 1993~GD and 2005~GL may be 
examples.
The time-averaged lightcurve-induced
diameter uncertainty is
%
$\sigma_{D}\langle t\rangle = \sigma_{D} \times S$,
%
where $S$ is a smoothing factor and 
is equal to $|\sin\phi|/\phi$, with $\phi=2\pi T/P$ (the rotational
phase, as above); $T$ giving the duration of the measurement;
and $P$ being the rotation period.

For all $T\gtrsim0.4~P$, it is the case
that $S\lesssim0.2$, so
$\sigma_D\langle t\rangle$ will almost always
be small for sufficiently long observations.
In cases where an asteroid's lightcurve
period is known, an observing plan that results
in small $\sigma_{D}$
can be created without requiring that the thermal and
reflected light observations be simultaneous or even phased.
Finally, we conclude that for very small asteroids, 
uncertainties introduced
by lightcurve effects will almost always be small, as follows.
Some very small asteroids have very short
rotation periods (just a few minutes), and most generally will require
long integration
times. Therefore, $T$ is likely to be $\gtrsim0.4~P$,
making $\sigma_{D}\langle t\rangle$ small.




\clearpage

\begin{deluxetable}{lcccccccc|rrrr|c}
\tabletypesize{\scriptsize}
\rotate
\tablecaption{Observing log \label{geom}}
\tablewidth{0pt}
\tablehead{
\colhead{Target} &
\colhead{AOR} &
\colhead{UT} & 
\colhead{$H$} & 
\colhead{$r$} &
\colhead{$\Delta$} & 
\colhead{$\alpha$} &
\colhead{t$_{{\rm frame}}$} &
\colhead{t$_{{\rm exp}}$} & 
\colhead{F$_{3.6}$} & 
\colhead{F$_{4.5}$} &
\colhead{F$_{5.8}$} &
\colhead{F$_{8.0}$} &
\colhead{Comment} \\
\colhead{} & 		
\colhead{} &		
\colhead{Date} &	
\colhead{(mag)} &	
\colhead{(AU)} &	
\colhead{(AU)} &	
\colhead{(deg)} &	
\colhead{(sec)} &	
\colhead{(sec)} &	
\colhead{($\mu$Jy)} &	
\colhead{($\mu$Jy)} &	
\colhead{($\mu$Jy)} &	
\colhead{($\mu$Jy)} &	
\colhead{}} 
\startdata
(6037) 1988 EG & 26984704 & 2008-Apr-07 21:34 & 18.7 & 1.24 & 0.42 & 46.96 & 12 & 120  & 50 (4) & 180 (7) & 856 (24) & 3013 (34) & 1 \\
1993 GD        & 26985216 & 2008-Apr-07 21:57 & 20.8 & 1.28 & 0.49 & 46.64 & 30 & 450  & 11 (3) & 36 (3)  & 124 (13) & 436 (13) & 2 \\
2005 GL        & 26984960 & 2008-Apr-07 18:35 & 21.2 & 1.37 & 0.71 & 44.06 & 100 & 1000 & \nodata & \nodata & 52 (7) & 108 (7) & 3 \\
\enddata
\tablecomments{
We list here the 
AOR (unique observation ID; these observations
were made as part of PID~476);
midtimes of the observations;
the Solar System absolute magnitude $H$, from the MPC;
the target heliocentric distance $r$,
Spitzer-centric distance $\Delta$, and
phase angle $\alpha$ at time of observation;
the individual frame time;
the total exposure time;
and the measured (not color-corrected) fluxes in the four 
IRAC bandpasses, with the errors in parentheses.
The effective wavelengths of these four bandpasses
are [3.550, 4.493, 5.731, 7.872]~$\mu$m.
The errors listed here do not include the
3\% absolute calibration uncertainty
\citep{reach05}.
It is difficult to estimate upper limit fluxes
at 3.6~and 4.5~$\mu$m for 2005~GL due to 
many faint star trails at the position of the asteroid.
Notes: (1) $H$ magnitude uncertainty around~0.15;
this object has a known lightcurve with period
near just under 3~hours and amplitude 0.2~mag.
(2) $H$ magnitude uncertainty around~0.4.
(3) $H$ magnitude uncertainty around~0.3.
}
\end{deluxetable}


\clearpage

\begin{deluxetable}{lcccc|cccl}
\rotate
\tablecaption{Physical properties of NEOs \label{results}}
\tablewidth{0pt}
\tablehead{
\colhead{Target} &
\multicolumn{4}{c}{Thermal fluxes ($\mu$Jy)} &
\colhead{diameter} & 
\colhead{albedo} & 
\colhead{$\eta$} & 
\colhead{Model} \\
\colhead{} & 	
\colhead{3.6~$\mu$m} &
\colhead{4.5~$\mu$m} &
\colhead{5.8~$\mu$m} &
\colhead{8.0~$\mu$m} &
\colhead{(m)} & 
\colhead{} &	
\colhead{} & 	
\colhead{}}
\startdata
(6037) 1988 EG & 10 & 142 & 806 & 2970 & 435 (23) & 0.31 (0.03) & 1.64 (0.11) & Floating $\eta$ \\ 
               & \nodata & \nodata & 806 & 2970 & 374 (34) & 0.43 (0.08) & 1.30 (0.18) & Floating $\eta$ \\ 
               & \nodata & \nodata & 806 & 2970 & 372 (35)  & 0.42 (0.07) & 1.29 (0.17) & Floating $\eta$, hot/cold MC proxy \\ 
               & \nodata & \nodata & 806 & 2970 & 414 (4) & 0.34 (0.006) & 1.52 & Fixed $\eta$ from $\alpha$ \\ 
               & 10 & 142 & \nodata & \nodata & 398 (15) & 0.37 (0.03) & 1.52 & Fixed $\eta$ from $\alpha$ \\ 
               & \nodata & \nodata & \nodata & \nodata & 399 (27) & 0.37 (0.05) & 1.45 (0.15) & Average of model results \\ \hline
%
1993 GD        & 6 & 31 & 117 & 430 & 143 (11) & 0.42 (0.07) & 1.02 (0.12) & Floating $\eta$ \\
               & \nodata & \nodata & 117 & 430 & 170 (31) & 0.32 (0.12) & 1.34 (0.36) & Floating $\eta$ \\ 
               & \nodata & \nodata & 117 & 430 & 184 (6) & 0.25 (0.02) & 1.52 & Fixed $\eta$ from $\alpha$ \\ 
               & 6 & 31 & \nodata & \nodata & 213 (17) & 0.19 (0.03) & 1.52 & Fixed $\eta$ from $\alpha$ \\ 
               & \nodata & \nodata & \nodata & \nodata & 178 (29) & 0.30 (0.10) & 1.35 (0.24) & Average of model results \\ \hline
%
2005 GL        & \nodata & \nodata & 49 & 107 & 147 (10) & 0.27 (0.03) & 1.48 & Fixed $\eta$ from $\alpha$ \\ 
               & \nodata & \nodata & \nodata & 107 & 145 (6) & 0.29 (0.02) & 1.48 & Fixed $\eta$ from $\alpha$ \\ 
               & \nodata & \nodata & 49 & \nodata & 191 (12) & 0.17 (0.03) & 1.48 & Fixed $\eta$ from $\alpha$ \\ 
               & \nodata & \nodata & \nodata & \nodata & 161 (26) & 0.24 (0.06) & 1.48 & Average of model results \\ 
\enddata
\tablecomments{
We list here the color-corrected thermal fluxes
(reflected light components subtracted) 
for each target.
Errors (omitted for clarity) on these thermal fluxes are
equal to the measurement errors given in Table~\ref{geom}
divided by
our derived color corrections of
[1.16, 1.09, 1.04, 1.01] for
[3.6, 4.5, 5.8, 8.0]~$\mu$m.
(The same color corrections apply for
all three targets. We neglect
uncertainties in reflected light flux, that is,
we assume that those uncertainties are zero).
We list solutions (with errors in parentheses)
to four-band and two-band sets
of data, indicating in the flux columns which 
measurements are being used. 
A range of models, discussed in the text,
are presented, as well as the average results
from the various models. The average results
are also plotted in Figure~\ref{sed}.
}
\end{deluxetable}

\clearpage



\clearpage






\end{document}